\documentclass[superscriptaddress,twocolumn,showpacs,preprintnumbers,
prb]{revtex4}
\usepackage{graphicx}
\usepackage{times}
\usepackage{epsfig}
\usepackage{color}
\usepackage{ulem}   
\begin{document}

\def\salto{\vskip 1cm} \def\lag{\langle} \def\rag{\rangle}

\newcommand{\redit}[1]{\textcolor{red}{#1}}
\newcommand{\blueit}[1]{\textcolor{blue}{#1}}
\newcommand{\magit}[1]{\textcolor{magenta}{#1}}

\newcommand{\MSTD} {Materials Science and Technology Division, Oak Ridge National Laboratory, Oak Ridge, TN 37831, USA}
\newcommand{\CNMS} {Center for Nanophase Materials Sciences, Oak Ridge National Laboratory, Oak Ridge, TN 37831, USA}

\title{Density-density functionals and  effective potentials
 in many-body electronic structure calculations}

\author{F. A. Reboredo}           \affiliation {\MSTD}
\author{P. R. C. Kent}        \affiliation {\CNMS}
\begin{abstract}
  We demonstrate the existence of different density-density
  functionals designed to retain selected properties of the
  many-body ground state in a non-interacting solution starting from
  the standard density functional theory ground state. We focus on diffusion quantum
  Monte Carlo applications that require trial wave functions with
  optimal Fermion nodes. The theory is extensible and
    can be used to understand current practices in several electronic
  structure methods within a generalized density functional framework.
  The theory justifies and stimulates the search of optimal empirical
  density functionals and effective potentials for accurate
  calculations of the properties of real materials, but also cautions
  on the limits of their applicability.  The concepts are
  tested and validated with a near-analytic model.
\end{abstract}

\today

\maketitle  
\section{Introduction}
Density Functional Theory (DFT) \cite{hohenberg,kohn} is based on the
Hohenberg-Kohn proof of a functional correspondence between the ground
state energy and the ground state density $E[\rho(r)]$.  In the
formulation of Kohn and Sham (K-S),\cite{kohn} the interacting
electron gas is replaced by non-interacting electrons moving in an
effective potential. In this construction, the non-interacting density
$\bar\rho(r)$ is equal to the interacting one $\rho(r)$ but no other
property of the interacting ground state is in principle retained in
the non-interacting wave function.  Initially DFT was formulated to
describe the total ground state energy of an interacting system and
$\rho(r)$.\cite{hohenberg,kohn,practice} Although progress towards
more accurate density functionals is ongoing, current approximations
such as the local density approximation (LDA)\cite{kohn,perdew81} and
more recent gradient-based extensions\cite{RMMartinBook} are already
successful in predicting many electronic properties of real
materials.  This success has led to the use of DFT beyond
its formal scope and unfortunately tempted some to believe that if we had the
exact ground state density functional, we would only need to solve
non-interacting problems even for properties not related to the ground
state energy and density.  While the virtues and limitations of the
Kohn-Sham {\it eigenvalues} are discussed in textbooks,\cite{RMMartinBook,parr}
the possible reasons for the success or failure of Kohn-Sham
{\it wave-functions} in many-body problems are little understood and widely
dispersed throughout the literature.

It is often assumed, without a formal proof, that the Kohn-Sham
non-interacting ground state wave-function forms a good description of
the interacting ground state wave-function to be used as the
foundation of theories that go beyond DFT such as GW-Bethe Salpeter
Equation (GW-BSE), Quantum Monte Carlo (QMC), or even configuration
interaction (CI).  This leads to an apparent contradiction in the
literature since density functionals that provide wave-functions that
are a good starting points in one field (as judged by comparison with
experiment) are found inadequate in others. Broadly summarizing: for
structural properties gradient corrected density functionals\cite{PBE}
are nowadays preferred over LDA.\cite{kohn,perdew81} Structural
properties depend essentially of atomic forces which in turn are
related to the density. However, for GW-BSE calculations of
  optical properties, an LDA-based ground state is
preferred\cite{aulbur00}. In this approach a good initial
approximation for the Green function is required.  In QMC calculations
a (non-interacting) Hartree-Fock (HF) ground state might be preferred
over LDA, but the subject is still under debate. In CI calculations,
instead, it is empirically claimed that the convergence with natural
orbitals\cite{lowdin55} is more rapid than HF orbitals.
  
In Diffusion Quantum Monte Carlo (DMC) a trial wave function enforces
the antisymmetry of the electronic many-body wave
function\cite{anderson79} and the nodal structure of the
  solution. The accuracy of the trial wave-function is critical and
determines the success or failure of the method to accurately predict
properties of real materials. The trial wave-function is usually a
product of a Slater determinant $\Phi_T({\bf R})$ and a Jastrow factor
$e^J({\bf R})$.  $\Phi_T({\bf R})$ is often constructed with single
particle Kohn-Sham orbitals or from other mean field approaches
  such as HF. The Jastrow, in turn, is a symmetric factor which does
not change the nodes, but accelerates convergence and improves the
algorithm's numerical stability.  The DMC algorithm finds the lowest
energy of the set of all wave-functions that share the nodes of
$\Psi_T({\bf R})$.  The exact ground state energy is obtained only if
the exact nodes are provided.  Since any change to an antisymmetric
wave-function must result in a higher energy than the antisymmetric
ground state, the energy obtained with arbitrary nodes is an upper
bound to the exact ground state energy.\cite{anderson79} Only in small
systems is it possible to improve the nodes
\cite{bajdich05,filippi00,umrigar07,rios06,luchow07} or even avoid the
trial wave-function approach altogether\cite{kalos00,zhang91}.
Consequently, a general formalism that could alleviate the nodal error
in large systems\cite{alfe04,reboredo05} is highly desired.  Quite
recently it has been shown that within the single Slater determinant
approach the computational cost of the DMC algorithm can have an
almost linear scaling with the number of
electrons\cite{williamson,alfe04,reboredo05,kolorenc}.  It is claimed,
if not formally proved, that the nodes of the many-body wave-function
are not too far from those of a wave function obtained via a mean
field approach.  However, this might not continue to hold as
electron-electron interactions become more
  important.  To improve the accuracy of these approaches and
increase the range of materials to which they can be applied it is
important to examine the advantages of different mean-field
wave-functions.

In this paper, we demonstrate that density-density functionals can be
obtained by finding the minimum of different cost functions relating
the set of non-interacting $v$-representable ground state with an
interacting many-body state.  The minimum of these cost functions
establishes a correspondence between the non-interacting and the
interacting wave-functions and their associated densities and
potentials.  The cost function can be designed to retain selected
properties of the many-body wave function in the non-interacting one. 
Crucially, for DMC applications the nodes can be optimized.
Under certain conditions density-density functionals exist that can
lead to standard scalar-density functionals.  As in the case of
standard DFT, this proof does not mean that we know the expression of
each functional or associated potential but it will certainly
stimulate the search of methods to find or approximate them.  For DMC
applications it is enough to prove that an optimal mean field
potential for nodes exists.  In order to test the theory, we find the
ground state wave function of a model interacting system.  Then we
obtain (i) the exact DFT effective exchange correlation potential
associated with the ground state density, (ii) the potential that
maximizes the projection of $\Phi_T({\bf R})$ with the ground state.
Finally, (iii) we optimize a potential to match the nodes and find that
surprisingly, for this model, the non-interacting solution in the same
potential as the interacting problem is a very good approximation for
the nodes while the exact non-interacting Kohn-Sham ground state is
particularly poor.

This paper is organized as follows: In Section II we
demonstrate the existence of an different density-density
correspondences associated with cost functions. We prove the existence
of this functional correspondence for the case of optimal nodes
required in DMC. In Section III we solve an interacting problem up to
numerical precision and find its many-body ground state wave-function.
Subsequently we optimize different cost functions to retain specific
properties of the ground state. Finally in Section IV we discuss the
relevance of our results for many-body electronic structure and give
our conclusions.

\section{Generalized density-density functionals}
Given an interaction in a many-body system, the Hohenberg-Kohn
theorem\cite{hohenberg} establishes a functional correspondence
between densities $\rho({\bf r})$, external potentials $V(r)[\rho({\bf r})]$ and ground state
wave-functions $\Psi({\bf R})[\rho({\bf r})]$;
where $[\rho({\bf r})]$ denotes a functional dependence on the
ground state density, and ${\bf R}$  denotes a point in the many-body $3 N$
space.  Since the density changes according to the strength and
functional form of the interaction, this correspondence is different
for different interactions. For a fixed interaction, the subset of
densities $\rho({\bf r}) $ corresponding to a ground state of an
interacting system under an external potential $V(r)$ are denoted as
{\it pure state} $v$-representable.\cite{parr} A non-interacting {\it
  pure state} $v$-representable density is given instead by
$\bar\rho({\bf r})= \sum_{\nu } |\phi _{\nu } \left({\bf r} \right)|^2
$ where $\phi _{\nu } \left({\bf r} \right)$ are the 
Kohn-Sham-like single particle orbitals, or eigenvectors, of the Hamiltonian:
\begin{equation}
\label{eq:Heff}
\left[ -\frac{1}{2}{\bf \nabla}^2 
+\bar V \left({\bf r}  \right) \right] \phi _{\nu }\left({\bf r} \right) 
=\varepsilon_{\nu } \phi _{\nu } \left( {\bf r}\right),
\end{equation}
where $\bar V \left({\bf r} \right) $ is an effective single particle
potential.  For simplicity we denote here a density to be
$v$-representable if it is both pure state {\it non-interacting} and
pure state v-representable. In the following we also imply
{\it pure state} when we write only $v$-representable.

\begin{figure}
\includegraphics[width=0.85\linewidth,clip=true]{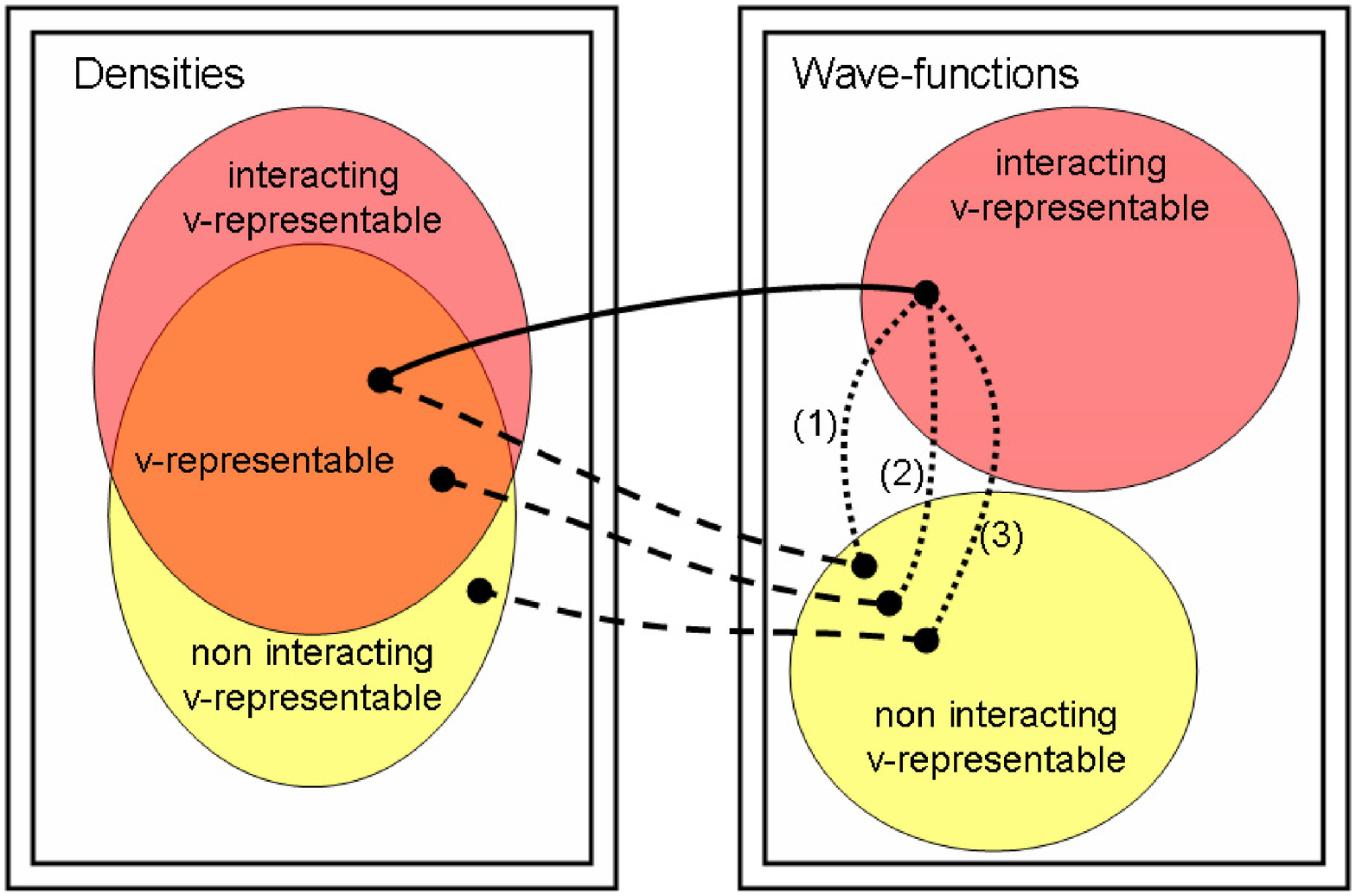}
\caption{ (Color online) 
  a) Representation of the sets of pure state $v$-representable
  interacting densities. b) Sets of interacting and non-interacting
  ground state wave-functions.  The Kohn-Sham formulation of DFT
  relates a $v-representable$ density with a pair interacting and a
  non-interacting wave-function. The same functional correspondence
  can be obtained minimizing Eq. (\ref{eq:Kr}) (line 1 in the figure).
  Different cost functions relate an interacting $v$-representable
  density with a different non-interacting-$v$ representable density
  (see lines 2 and 3).
\label{fg:scheme}}
\end{figure}

Each point in the sets of $v$-representable densities is associated
with two {\it different} points in the wave-functions Hilbert space.
In figure \ref{fg:scheme} we schematize the subset of
$v$-representable densities and the functional correspondence with the
subsets of the interacting and non-interacting ground state
wave-functions. Note that, in principle, the two subsets of ground
state wave-functions do not necessarily overlap. In the non
interacting case the wave-function is given by a Slater determinant of
Kohn-Sham-like orbitals but for interacting problems this simplification
is not longer possible.

The Kohn-Sham scheme for density functional theory establishes a
correspondence between interacting and non-interacting wave-functions
represented as line (1) in Fig \ref{fg:scheme}. This Kohn-Sham
correspondence between wave-functions is implicit in the Khon-Sham
construction for the external effective potential.\cite{kohn} Figure
\ref{fg:scheme} emphasizes that the wave-functions joined by line (1),
while different, give the same electronic density. In more technical
terms they both belong to the same Percus-Levy partition of the
Hilbert space\cite{percus,levy} but they are the minimum energy
wave-function for different interactions.  The exchange-correlation
potential is by construction the {\it difference} that one has to add
to the external potential in a non-interacting problem so that its
ground state density is the same as the interacting one.  If the
energy-density functional $E[\rho({\bf r})]$ is known, the effective
non-interacting potential can be obtained following the standard
Kohn-Sham approach. If the ground state density $\rho({\bf r})$ is
known, the same correspondence between interacting and non-interacting
densities can be achieved by minimizing the following function
\begin{equation}
\label{eq:Kr}
K_{\rho}= \frac{1}{2} \int {\bf dr} 
\left[ \bar \rho({\bf r})- \rho({\bf r}) \right]^2.
\end{equation}
within the subset of non-interacting $v$-representable densities.
Formally, this could be done by exploring all values $\bar V\left({\bf
    r} \right) $ in Eq (\ref{eq:Heff}).  

In practice, if the density of the interacting ground state is known,
the potential $\bar V_{K_{\rho}}({\bf r})$ that minimizes Eq.
(\ref{eq:Kr}) can be obtained numerically with a procedure similar in
spirit to the optimized effective potential (OEP) method.  The change
in the density required to minimize Eq.  (\ref{eq:Kr}) is
\begin{equation}
\Delta_{\rho} = - [\bar \rho({\bf r})- \rho({\bf r})] .
\end{equation}
Within linear response, the change in the {\it potential} required to
produce $\Delta_{\rho}$ is
\begin{equation}
\label{eq:vKr}
\Delta \bar V_{K_{\rho}}({\bf r})  = 
    \int  {\bf dr^{\prime} } 
\left[ \rho({\bf r ^{\prime} })-\bar \rho({\bf r^{\prime} }) \right] 
\frac{\delta V \left( {\bf r ^{\prime} } \right) }{\delta \rho\left( {\bf r} \right) } 
\end{equation}
Adding recursively $\Delta \bar V_{K_{\rho}}({\bf r})$ we can find the
potential $\bar V_{K_{\rho}}({\bf r})$ associated with $K_{\rho} = 0$
(see an example below).  

\subsection{Other density-density correspondences}
\label{lb:df}
It is often desirable o preserve properties in addition to the density
of the many-body ground state $\Psi({\bf R})$ in a non-interacting
wave-function $\Phi_{T}({\bf R})$ to be used as a starting point for
theories that go beyond DFT.  This task involves exploring all the
non-interacting $v$-representable set in order to find a wave-function
that best describes a given property. This is a typical optimization
problem.  One of the most common strategies in optimization is the
design of a cost function.
One example is Eq (\ref{eq:Kr}), a measure of the difference in
two densities. Another example of a cost function is
\begin{equation}
\label{eq:KDet}
K_{Det}=- \left| \left< \Psi\right. \left|\Phi_{T}\right> \right|^2
\end{equation}
which involves a projection of the interacting ground state $\Psi$ in
the set of non-interacting $v$-representable wave-functions
$\{\Phi_{T}\}$.  The minimum of Eq. (\ref{eq:KDet}) is the non-interacting
ground state Slater determinant with maximum projection in the
interacting ground state.  We have claimed above that the interacting
wave-functions might be in general very different from a single
non-interacting Slater determinant.  Accordingly, we expect $K_{Det} >
-1 $.

We expect to find a different minimum in the non-interacting
ground state set, if we change the functional form of the cost
function from Eq. (\ref{eq:Kr}) to Eq. (\ref{eq:KDet}) for the
following reasons:

1) We can visualize the cost function as a scalar potential defined in
the full Hilbert-space. Although different cost functions can share
the same minimum in the complete Hilbert space, in the restricted
subset of non-interacting ground state wave-functions, different cost
functions can have a different minimum: the optimal point found
depends on the functional form of the cost function.  Accordingly,
while all the cost functions we propose here would be minimized if we
could reach the interacting many body state $\Psi$ ( where, of course,
every property is retained exactly), because our search is constrained
to non-interacting $v$-representable subset the minimum we would
find will depend on the properties we wish to retain.

2) The Hohenberg Kohn theorem, when applied to the non-interacting
$v$-representable case implies that, in the absence of degeneracy, 
there is at most {\it one} wave-function that has the same density as the interacting case.
Therefore, once the minimum of an {\it arbitrary} cost function is found,
its associated non-interacting density can no longer be equal to
the interacting density unless the property enforced by the cost
function can be related back to density. Enforcing the
non-interacting density to remain equal to the interacting ground
state density prevents all other properties of the non-interacting
wave-function from being further improved. If we intend to optimize other
properties, we have to relax the density constraint finding a
different wave-function associated with a different density.

The minimization of different cost-functions, relating the interacting
ground state $\Psi({\bf R})$ with the non-interacting
$v$-representable set, provide in-principle different correspondences
between interacting and non-interacting wave-functions represented as
different lines in figure \ref{fg:scheme}. Each cost function $K$
defines a correspondence different than the identity between pure
state $v$-representable densities and non-interacting pure-state
$v-$representable densities.  As a consequence, the idealized
optimization processes outlined here defines an operator $U_K$ that
turns each $\rho({\bf r})$ into a non-interacting density
corresponding to the wave-functions $\Phi_T({\bf R})$ which is the
minimum of a cost function $K$.
\begin{equation}
\label{eq:U}
\bar\rho_{K}({\bf r}) = U_{K}\left[\rho({\bf r})\right].
\end{equation} 
Note that if the minimum of a given cost function $K$ is a single
$\Phi_T({\bf R})$ for every $v$-representable density, then $ U_{K}$
defines a density-density functional. When more than one
non-interacting $v$-representable wave-function give the same optimal
value for $K$, the degeneracy can be broken by additional requirements
in the cost function [e.g. also minimizing Eq.  (\ref{eq:Kr}), the difference
between the current and pure state densities ].  Since we only need
one optimal wave-function, any from a degenerate minimum can be chosen
to construct the density-density functional $U_{K}$.  When
minimization of a cost function defines a one to one correspondence
with an inverse, a more usual energy-density functional of the form
$E\{ U_{K}^{-1}[\bar\rho_{K}({\bf r}) ]\}$ can be constructed. Only a
restricted class of cost functions lead to density
transformations with an inverse. Minimization of the cost functions
among all pure-state-non-interacting  $v$-representable densities
defines the optimal effective potential which is a function of this
density.

Given a cost function, $K$, finding an approximation for the density
transformation operator $U_{K}$ could certainly be as demanding as
finding an approximation for the energy-density functional
$E\left[\rho({\bf r})\right]$ required by DFT based methods.  This
task is beyond the goal of this paper. However, we will show that we
can expect the operator associated to the best nodes for DMC
($U_{DMC}$) to be non-local and very different from the identity.
Accordingly we can expect non-interacting wave-functions with good
nodes to be a poor source of densities. Moreover, for the example
considered below, we find, that the direction we might have to explore
to optimize the potential might be surprisingly different than the
attempts considered so far\cite{wagner03,kolorenc}.

\subsection{The Diffusion Monte Carlo case}
We next show that optimization of the nodes for DMC among the set of
$v-$representable wave-functions leads to a correspondence between
{\it pure state} $v$-representable densities and  pure state
non-interacting $v$-representable densities of the class described
above. These in turn demonstrate the existence of an optimal effective
non-interacting nodal potential.

Since, the ground state density $\rho({\bf r})$ determines the ground
state wave-function $\Psi({\bf R})[\rho({\bf r})]$,\cite{hohenberg}
$\rho({\bf r})$ defines also the points ${\bf R}$ of the nodal surface
$S_0({\bf R})[\rho({\bf r})]$ where $\Psi({\bf R})[\rho({\bf r})] =
0$. We can also classify the nodal surfaces in pure state v-representable
and pure-state-non-interacting $v$-representable.

The DMC algorithm in the fixed node approximation finds the lowest
energy of the set of all wave-functions that share the nodes or the
trial wave-function. For Slater determinant Jastrow wave-functions,
the nodes of the trial wave-function are by construction those of
$\Phi_T({\bf R}) $; that is they are pure-state non-interacting
$v$-representable. The DMC energy, $E_{DMC}$ is also a function of the
external potential which in turn is a function of the interacting
ground state density $V(r)[\rho({\bf r})] $.  Thus minimization of
$E_{DMC}[\Phi_T({\bf R}),\rho({\bf r})]$ in the set of non-interacting
$v$-representable wave-functions $\Phi_T({\bf R})$ determines one
$\Phi_T({\bf R})$ with the best nodes. Every optimal $\Phi_T({\bf R})$
defines an optimal auxiliary density $\bar\rho_{DMC}({\bf r})$. As a
consequence optimizing the nodes of the trial wave-function by
perturbing the nodes of pure state non-interacting wave-functions
implies finding another correspondence between interacting and
non-interacting densities (another line in figure
\ref{fg:scheme}). The best cost function for optimal nodes is
ultimately the DMC energy itself.

Since we restrict the search to pure-state non-interacting
$v$-representable nodes, the minimum energy $E_{DMC}[\rho({\bf r}]$
will be larger than the true ground state energy $E[\rho({\bf r}]$,
because of the upper bound theorem, unless $S_0({\bf R})$ is
non-interacting $v$-representable.

Note that for an arbitrary interaction $S_0({\bf R})$ is not expected
to be, in general, pure-state-not-interacting v-representable.
However, if $S_0({\bf R})$ were non-interacting v-representable, the
best Slater Determinant $\Phi_{T}({\bf R})$ for DMC could be formally
found by finding the minimum of the cost function
\begin{equation}
\label{eq:Ks0}
K_{S_0}=\int_{S_0} {\bf dS} \left| \Phi_{T}({\bf R}) \right|^2.
\end{equation}
where $\int_{S_0}$ denotes a surface integral over the interacting nodal
surface.

\section{Cost function minimization}

To demonstrate the theoretical concepts above we solve a
simple non-trivial interacting model as a function of the interacting
potential strength and shape. We then optimize the wave-functions to
minimize the cost functions in Eqs.  (\ref{eq:Kr}), (\ref{eq:KDet})
and (\ref{eq:Ks0}) so as to find the exact DFT wave-function, the
wave-functions that maximize the projection on the interacting ground
state and minimize the projection on the nodes. Subsequently, we
estimate the volume of the Hilbert space enclosed between the
nodes of the interacting wave-functions and the optimized
non-interacting ones.

\subsection{A model interacting ground state}
For illustrative purposes we choose the interacting problem to be as
simple as possible and yet not trivial. We solve the ground state of
two spin-less electrons moving in a two dimensional square of side
length 1 with a repulsive interaction potential of the form $V({\bf
  r},{\bf r^{\prime}}) = 8 \gamma \cos{[\alpha
  \pi(x-x^{\prime})]}\cos{[\alpha \pi(y-y^{\prime})]}$. \cite{units}
While this potential is different than the Coulomb interaction, it
shares some of its properties. For positive $\gamma$ and $| \alpha |
< 1$ the interaction is repulsive with a repulsion that increases
monotonically when for shorter distances. The amplitude of the
repulsion as compared to the kinetic energy can be changed by
adjusting $\gamma$. Since the Coulomb interaction is self similar,
changing $\gamma$ mimics what happens in a real system when we
change the size of the system. The shape of the potential within the
confined region can be altered by changing $\alpha$. In the limit of
$\alpha \rightarrow 0$ the interaction potential is separable which
allows several limits to be tested (such as the nodes). The functional
form facilitates an analytical treatment of the problem by
removing the singularity of the Coulomb interaction at short
distances.

We expanded the many-body wave-function in a full CI on non-interacting
Slater determinants with the same symmetry as the
ground state. The ground state is degenerate because there are only
two electrons.  We chose one of the ground state wave-functions
according to the $D_2$ subgroup of the $D_4$ symmetry of the
Hamiltonian. With this choice, $ \rho({\bf r}) $ has $D_2$ symmetry
($x$ is not equivalent to $y$). The basis of Slater determinants was
constructed with functions of the form $16 \mathrm{sin}( n \pi x) \mathrm{sin}( m \pi y)
$ with $n$ and $m < 8$. Since parity is preserved by the interaction
and Slater determinants of identical functions are zero, the size
of the basis set is reduced to only 300 in our calculations.
 
Most of the calculations reported here were done analytically with the
help of the Mathematica package, including all of the
electron-electron interaction integrals.\cite{notebook} The only
source of errors are numerical truncation and the size of the basis,
which was tested for convergence.

\begin{figure}
\includegraphics[width=0.85\linewidth,clip=true]{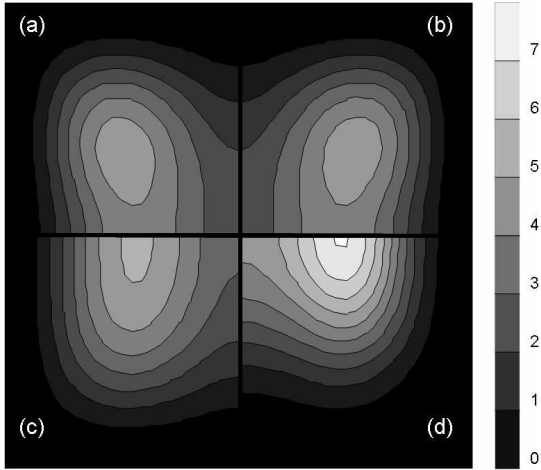}
\caption{Ground state densities, in particles per unit area, for 
  two interacting spin-less electrons in a square box. The complete
  density can be obtained by reflections (see also Fig
  \protect\ref{fg:potentials}). (a) Full CI ground state interacting
  density. (b) Exact DFT solution obtained minimizing Eq.
  (\ref{eq:Kr}).  (c) Slater determinant with maximum projection with CI
  the ground state [see Eq.  (\ref{eq:KDet})].  (d) Slater determinant
  with minimum amplitude on the nodes of the CI ground state [see Eq.
  (\ref{eq:Ks0})].
\label{fg:densities}}
\end{figure}

In Figure \ref{fg:densities} we show the quadrants of densities
corresponding to wave-functions that are even for reflections in the
$y$ direction and odd in the $x$ direction for $ \gamma = 2$ and
$\alpha = 1$. Figure \ref{fg:densities}(a) shows the upper left
quadrant of the density of the interacting ground state of two
spin-less electrons obtained with full CI.  Figure
\ref{fg:densities}(b) shows the upper right quadrant of the
non-interacting density corresponding to the effective potential
obtained by adding recursively $\Delta V_{K_{\rho}}({\bf r}) $ [Eq.
(\ref{eq:vKr})], which is exactly the reflection of $\rho({\bf
  r^{\prime} })$ Fig.  \ref{fg:densities}(a) up to numerical precision
[$K_{\rho} = 0 $ in Eq.  (\ref{eq:Kr})].  Because of the
  Hohenberg-Kohn theorem,\cite{hohenberg} $ V_{K_{\rho}} $ and the
  Kohn-Sham potential $V_{KS}(\rho) $ can only differ by a constant
  and thus the wave-functions coming from this potential are the exact
  DFT wave-functions for our interaction. The properties of the
wave-functions will be discussed later in the text.  The densities in
Figs.  \ref{fg:densities}(d) and \ref{fg:densities}(d) correspond to
the minimum of the cost functions given in Eqs (\ref{eq:KDet}) and
(\ref{eq:Ks0}) obtained as described below.

\subsection{Effective potential optimization}
We now consider more difficult cost functions than density
differences, Eq. (\ref{eq:Kr}).  For non-interacting $v$-representable
densities there are also functional correspondences between ground
state wave-functions, potentials and densities. This concept has been
exploited in the optimized effective potential (OEP) for exact
exchange.  \cite{xxtheory,xxsemiconductors,xx2deg,us} The exchange
potential can be calculated in OEP
\cite{xxtheory,xxsemiconductors,xx2deg,us} as:
\begin{eqnarray}
\label{eq:vXX}
&& V_{x}({\bf r} ) = \frac{\delta E_{x}}{\delta \rho\left({\bf r} \right) 
} \\
&& =  \sum_{\nu }^{occ} \int \!\!\!\!\int \!\!{\bf dr^{\prime }} {\bf dr^{\prime \prime }} \left[ 
\frac{\delta E_{x}}{\delta \phi _{\nu } \left( {\bf r^{\prime \prime}
}\right) }\frac{\delta \phi _{\nu } \left( {\bf r^{\prime \prime} }\right) 
}{\delta V_{KS}\left( {\bf r^{\prime }} \right) }+c.c.\right] \!\! 
 \frac{\delta V_{KS}\left( {\bf r^{\prime} } \right) }{\delta
\rho\left( {\bf r}  \right) }.  \nonumber
\end{eqnarray}
In Eq. (\ref{eq:vXX}), the functional derivative $\delta E_{x}/\delta
\phi _{\nu } \left( {\bf r} \right) $ is evaluated directly from the
explicit expression for the exchange energy $E_{x}$ in terms of $\phi
_{\nu } \left( {\bf r }\right) $. Next $\delta \phi _{\nu } \left(
  {\bf r} \right) /\delta V_{KS}\left( {\bf r^{\prime }} \right) $ is
evaluated using first-order perturbation theory from Eq.
(\ref{eq:Heff}). Finally $ {\delta V_{KS}\left( {\bf r^{\prime} }
  \right) }/ {\delta \rho\left( {\bf r} \right) } $ is the inverse of
the linear susceptibility operator. If there are fixed boundary
conditions such as the number of particles, the susceptibility
operator is singular \cite{xxsemiconductors,us}. Excluding these null
spaces it can be inverted numerically.\cite{xxsemiconductors} We use
earlier this susceptibility operator in Eq. (\ref{eq:vKr}). Equation
(\ref{eq:vXX}) is by construction the gradient of the exchange energy
in the set of pure-state-non-interacting $v$-representable
densities. While we are not going to attempt an exact exchange
approach in this paper, the ability to calculate gradients allow us to
minimize cost functions as long as the cost function $K$ can be
expressed in terms of non-interacting ground state wave wave-functions
or eigenvalues. The potential that minimizes $K$ can be obtained by
recursively applying the formula
\begin{equation}
\label{eq:vK}
\delta V_{K}({\bf r}  )  =  
\epsilon \sum_{\nu }^{occ}\int \!\!
 {\bf dr^{\prime }} \!\!
\frac{\delta K} {\delta \phi _{\nu } 
\left( {\bf r^{\prime } }
\right) 
}
\frac{\delta \phi _{\nu } \left( {\bf r^{\prime } }\right) }
{\delta V_{KS}\left( {\bf r }  \right) }+c.c.
\end{equation}
Equation (\ref{eq:vK}) gives the direction we need to change the
potential to minimize the cost function. The magnitude of the change
is controlled by $\epsilon$, which can be adjusted as one reaches the
minimum.

Replacing $K$ by $K_{Det}$ in Eq. (\ref{eq:vK}) and using Eq
(\ref{eq:KDet}) and first order perturbation theory we find
 \begin{eqnarray}
\label{eq:vKDet}
\delta V_{K_{Det}}({\bf r}  )  &=&  
\epsilon \left< \Psi\right. \left|\Phi_{T}\right>
\sum_{\nu }^{o}
\sum_{n}^{u} 
\left< \Psi\right| c^{\dag}_n c_{\nu} \left|\Phi_{T}\right>
\frac{\phi_{n}({\bf r }) \phi_{nu}({\bf r })}
     {\varepsilon_{\nu}-\varepsilon_n} \nonumber \\ 
&+&c.c.
\end{eqnarray}
In equation (\ref{eq:vKDet}) $\sum_{n}^{o}$ ( $\sum_{n}^{u}$ ) means
sum over occupied (unoccupied) states, while $c^{\dag}_{n}$ and
$c_{\nu}$ are creation and destruction operators on the
non-interacting ground state $\left|\Phi_{T}\right>$. One can
understand also the state $c^{\dag}_n c_{\nu} \left|\Phi_{T}\right>$
as the many body wave-function $\Phi_{T}^{n,\nu}(R)$ resulting from
replacing the occupied state $\phi_{\nu}$ by the $\phi_{n}$. This is
equivalent to creating an electron hole pair excitation in a
non-interacting ground state.  In Eq. (\ref{eq:vKDet}) a term in the
potential is added every time an electron hole pair excitation has no
zero projection to the interacting ground state.  Since the basis of
products of wave-functions $\phi_{n}({\bf r }) \phi_{\nu}({\bf r }) $
is over-complete, there are linear combinations with non-zero
coefficients that add up to zero.  A minimum is found when the
gradient of the cost function with respect to variations of the
effective potential is zero. If the absolute minimum is found, the
wave function can only be improved further by a multi-determinant
expansion, that is, outside the set of pure-state non-interacting
densities.  Since we choose a basis expansion for the single particle
orbitals to be sine functions, the products $\phi_{n}({\bf r })
\phi_{nu}({\bf r }) $ are linear combinations of sine products.  These
sine products can be transformed analytically to cosines.  The change
in the potential is thus written in a cosine basis which is complete.
All coefficients must vanish in the cosine basis when a minimum is
found. This allows us to verify that the gradient in the potential can
be minimized up to numerical precision.  The integrated effective
potential is thus naturally expressed as a linear combinations of
cosines, which allows the analytical calculation of the coefficients
of the effective potential matrix in a basis of sines, where the
kinetic energy is diagonal. The Slater determinant $|\Phi_T>$ is
written in the same basis as the interacting ground state $|\Phi>$.
The projections involved in Eq.  (\ref{eq:vKDet}) are then reduced to
a scalar product of the vectors of coefficients.

In figure \ref{fg:densities}(c) we show the ground state density
associated to minimization of Eq. (\ref{eq:KDet}) for the same
parameters as the interacting ground state density in Fig
\ref{fg:densities}(a). We see that while optimizing the cost function
(\ref{eq:Kr}) allows matching the interacting density exactly,
optimizing the wave-function projection requires a significant change in
the resulting density.

 Similarly, replacing $K$ by $K_{S_0}$ in Eq. (\ref{eq:vK}) and using Eq (\ref{eq:Ks0}) 
we get
\begin{eqnarray}
\label{eq:vKs0}
\delta V_{K_{S_0}}({\bf r}  )  &=&  
\epsilon 
\sum_{\nu }^{o}
\sum_{n}^{u}
 \int_{S_0}\;\; {\bf dS}  \Phi_{T}^{n,\nu}({\bf R}) \Phi_{T}({\bf R}) 
\frac{\phi_{n}({\bf r }) \phi_{nu}({\bf r })}
     {\varepsilon_{\nu}-\varepsilon_n} \nonumber \\
&+& c.c.
\end{eqnarray}
Unlike Eqs.\ref{eq:vKr} and \ref{eq:vKDet}, a complication appears when evaluating the
integral over the nodal surface $ \int_{S_0} {\bf dS} $. This integral involves
finding the points where the many-body wave-function is zero. The
problem is simplified because the derivatives of the many body
wave-functions can be obtained analytically. Consequently, starting
from an arbitrary point ${\bf R}$ we can find a zero recursively with
the Newton-Raphson method, $ {\bf R}_{n+1}= {\bf R}_{n} + \Psi({\bf
  R}) {\bf \nabla} \Psi({\bf R}) / \left| {\bf \nabla} \Psi({\bf R})
\right |^2 $. Next we make a random displacement ${\bf \Delta R}$ in
the hyper-plane perpendicular to ${\bf \nabla} \Psi({\bf R})$ within a
circle of radius 0.05 and find a node again.  We repeat this process
$50$ times and select an element of $\{{\bf R} \}_S$. With this
parameters, the random position ${\bf R}_n$ can travel across the full
size of the system so that the distribution is homogeneous.  By
repeating this process $N=500$ times, excluding points at the
boundaries which are zero by construction, we generate an homogeneous
distribution of points at the nodal surface $\{ {\bf R}\}_S$. We
approximate the integral in Eq. (\ref{eq:Ks0}) as a sum on the values
on the set $\{ {\bf R}\}_S$.  Note that while the total area of the
surface would be in general involved as a factor, the value of this
area is not relevant since we are interested in finding a minimum of
the cost function and the position of the minimum of any function is
not altered by a positive multiplicative constant. The sum over random
points introduces a relative error of order $1/\sqrt{500}$ . Replacing
the integral with a summation creates also many local minima in the
landscape of Eq. (\ref{eq:vKs0}). Accordingly, we tested different
initial conditions; the best results are obtained starting from
$V_{K_{S_0}} = 0$.

The density
resulting from minimization of Eq. (\ref{eq:Ks0}) is plotted in Fig
\ref{fg:densities}(d). We see here again a significant change as
compared with the fully interacting CI ground state [see Fig
\ref{fg:densities}(a)] and the the exact DFT non-interacting solution
Fig \ref{fg:densities}(b).

Figure \ref{fg:densities} is a clear example that corroborates our
claim in Section \ref{lb:df} that enforcing different properties on the
non-interacting wave-function implies a density-density correspondence
different than the identity between the interacting and non
interacting systems. Similar results are observed as function of the
strength and shape of the interaction [controlled by $\gamma$ and  $\alpha$]  
 
A comparison between Eqs. (\ref{eq:vKDet}) and (\ref{eq:vKs0}) clearly
shows that the relative values of the coefficient multiplying
$\phi_{n}({\bf r }) \phi_{nu}({\bf r }) $ depends fundamentally on the
cost function.  Therefore, even starting from the same effective
potential and $\Phi_{T}({\bf R})$ the coefficient affecting each
individual product $\phi_{n}({\bf r }) \phi_{nu}({\bf r }) $ depends
on the functional form of the cost function.  This change in the 
potential remains present when the potential is written in the complete
cosine basis.  Thus, the effective potential must change in accord
with the property of the interacting ground state that one aims to
enforce in the non-interacting ground state with a cost function.

\begin{figure}[b]
\includegraphics[width=0.85\linewidth,clip=true]{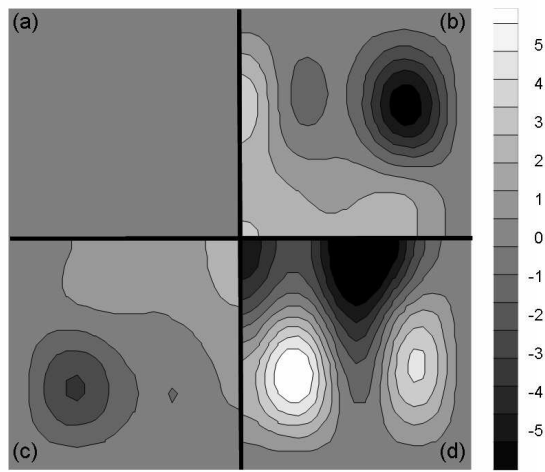}
\caption{Optimized effective potentials corresponding to the densities 
  in Fig. \protect\ref{fg:densities}. The complete potentials can be
  obtained by reflection on the black lines. Gray level
  values\cite{units} are given on the right. The optimal effective
  potentials are strongly dependent on the property we target to retain
  in the wave-function.}
\label{fg:potentials}
\end{figure}

Figure \ref{fg:potentials} shows the effective potentials used for the
calculations shown in Fig \ref{fg:densities}. We show in Fig
\ref{fg:potentials}(a) a constant, since in the interacting problem
solved with full CI no effective external potential was added. Figures
\ref{fg:potentials}(b) [minimum of Eq. (\ref{eq:Kr})],
\ref{fg:potentials}(c) [minimum of Eq. (\ref{eq:KDet})] and
\ref{fg:potentials}(d) [minimum of Eq. (\ref{eq:Ks0})] show a clear
change in the effective potential depending on the cost functions. 
As argued earlier  Fig. \ref{fg:potentials}(b) shows the exact Kohn-Sham DFT
potential for this interaction which implies that a different density 
functional must be used to obtain non-interacting wave-functions preserving
properties other than the density. 

Note that Eq. (\ref{eq:Ks0}) could be zero only for a
pure-state-non-interacting $v$-representable nodal surface.  However,
if the nodes are not, replacing in $K_{S_0}$ could result in a
potential that simply prevents the non-interacting wave-function to
reach regions of space where the nodes are more troublesome. The
potential shown in Fig. \ref{fg:potentials}(d) presents a maximum in
regions where instead Figs. \ref{fg:potentials}(b) and
\ref{fg:potentials}(c) develop a minimum. These are the regions where
the electrons in the many-body wave-functions tend to localize
because of correlation effects. The maximum in Fig
\ref{fg:potentials}(d) suggest the possibility of non-$v$
representability by a non-interacting wave-function in this model.

\subsection{Wave-function internal structure}

In order to quantitatively test the quality of the nodes of the
wave-functions found by minimization of Eqs. (\ref{eq:Kr}),
(\ref{eq:KDet}) and (\ref{eq:Ks0}) and to test the convergence of the
nodes of the full CI calculations, we take advantage of the
homogeneous distribution of points $\{{\bf R} \}_{S}$ at the nodal
surface $S_0({\bf R})$ described earlier.

For each point ${\bf R}$ in $\{{\bf R} \}_{S}$ we can find the
distance $\ell_i$ to the node of another wave-function $\Phi_T({\bf
  R})$ in the direction of $ {\bf \nabla} \Psi({\bf R})$.  Thus
\begin{equation}
\label{eq:DV}
\Delta V = \frac{1}{N} \sum_i \ell_i 
\end{equation}
is an approximated measure of the fraction of the Hilbert space volume between the nodal surface 
of $\Phi_{T}({\bf R})$ and $\Psi({\bf R})$  and 
\begin{equation}
\label{eq:dr}
\delta \rho = \frac{1}{3 N} \sum_i \ell_i \left|\Phi_T \left({\bf R}_i
  \right) \right| ^2 
\end{equation} 
measures the probability density inside $\Delta V$. 

We can use Eqs. (\ref{eq:DV}) and (\ref{eq:dr}) to test the
convergence of the CI ground state nodes as a function of the size of
the basis set. While the ground state energy requires $40$ basis
functions, the nodes are more difficult to converge requiring
four times as many. The size of the basis required to converge the nodes
was determined for $\gamma = 2$ plotting $\delta \rho$ between the CI
ground state with 300 wave-functions and the CI ground state obtained
using a reduced basis.

\begin{figure}
  \includegraphics[width=0.95\linewidth,clip=true]{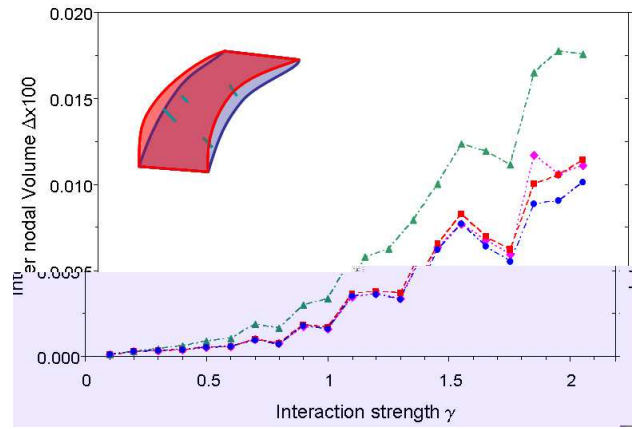}
\caption{(Color online)
  Fraction of the Hilbert space $\Delta V $ between the full CI node
  and the nodes of different optimized wave-functions.  Triangles
  correspond to the exact DFT wave-function [Eq.
  (\ref{eq:Kr})], squares to maximum projection [Eq.  (\ref{eq:KDet})],
  rhombi to the minimum amplitude at the nodes [Eq.  (\ref{eq:Ks0})]
  and circles to the non-interacting ground state. The inset shows the
  method used to estimate $\Delta V$
\label{fg:volumes}}
\end{figure}

The quantities in Eqs. (\ref{eq:DV}) and (\ref{eq:dr}) can be used
also to characterize the nodes of different wave-functions as compared
with the exact node.  In figure \ref{fg:volumes} we show the volume
enclosed between the nodes $\Delta V$ of different optimized
$\Phi_{T}({\bf R})$ and the interacting ground state $\Psi({\bf R})$
as a function of the strength of the interaction potential $\gamma$.
Note that the Kohn-Sham DFT solution gives a significantly larger
volume than other optimized wave-functions. The difference increases
as the interaction strength increases.  The wave-function that results
from minimizing Eq. (\ref{eq:KDet}) which targets wave-function projection
fares very well over the range explored. In turn, minimization of
$K_{S_0}$ [see Eq.  (\ref{eq:Ks0})] results in nodes that are only
sometimes marginally better.  Surprisingly, the non-interacting
solution, that is the non-interacting ground state in the {\it
  absence} of any effective potential, is remarkably good. Similar
results are found by altering the shape of the potential with
$\alpha$.

\begin{figure}
\includegraphics[width=0.95\linewidth,clip=true]{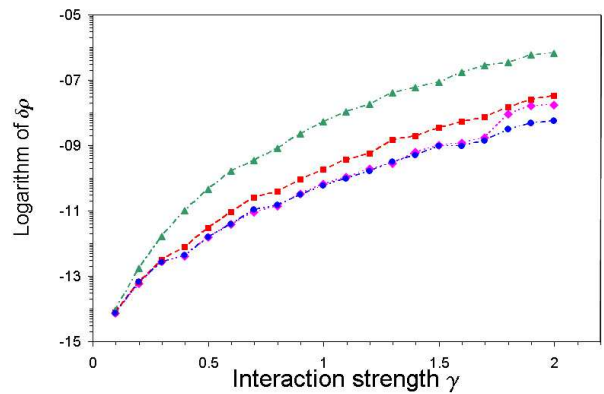}
\caption{(Color online)
Probability density inside the volume between the
nodes of the full CI wave-function and optimized wave-functions.
Same conventions and symbols as in Fig. \protect\ref{fg:densities}}
\label{fg:volden}
\end{figure}

In figure \ref{fg:volden} we plot the values $\delta \rho $ for
different optimized wave-functions as a function of $\gamma $. We see
again that the exact Kohn-Sham DFT solution is not the best. The quantity
$\delta \rho $ is a measure on how much the error in the nodes would
affect the probability density and thus it can be understood as the as
a measure of the nodal error in the ground state energy. Again
in this case the non-interacting ground state without any effective
potential is the best approximation. 

\section{Discussion}
Although the numerical investigation of the different density-density
functionals described above required numerical representation of the
many body ground state wave-function, the conclusions that we draw have
general value. The model we explore is simplified but has the
advantages that the results can be converged and are free of significant
approximations.  The simplified interaction used in the model
  retains essential features of the Coulomb interaction.

We have shown (Fig. \ref{fg:densities} and \ref{fg:potentials}) that
the effective densities and potentials are explicit functions of a
cost function.  Potentials and densities very different to the exact
DFT solutions are obtained if we enforce properties beyond $\rho({\bf
  r})$ in the cost function. The exact DFT wave-function matches
$\rho({\bf r})$ with {\it complete} disregard to other elements of the
many body wave-function structure.  Since the Hohenberg-Kohn
theorem\cite{hohenberg} is valid, optimizing other properties of the
non-interacting ground state in general requires changing the
potential with a resulting impact on the density.

We find that mean field methods, while giving an accurate description
of the density can mislead us in other aspects of the wave-function
structure such as the nodal surface. In this paper we argue that,
among the pure-state-non-interacting $v$-representable densities,
there is at least one that more accurately describes the interacting
ground state nodes.  We can optimize the wave-function associated with
this density with a cost function. The cost function form depends on the
property we target to retain and also the optimal density we find. For
the nodes, the optimal cost function is clearly the DMC energy.  A
fixed cost function establishes a density-density correspondence which
can be described as an operator $U$ that transforms interacting
densities into non-interacting ones.

While finding the functional form of $U_{DMC}$ is a task beyond the
scope of this paper, we argue that we can expect this operator to be
highly non-local and very different from the identity, in particular
for strong electron-electron correlations. We find that we can improve
the nodes with some simple cost functions, but the best nodes we found
were obtained solving the non-interacting problem without the addition
of any effective potential. We cannot exclude the possibility that
this result might well be an accident of the model. Our result, however,
shows that the popular expectation that the DFT solution is a good
starting point for nodes is not valid in general.

Optimal wave-functions can be found by altering an external potential.
This idea is not new. In practice wave-functions are optimized with
the trial wave-function only minimizing the ground state energy
$K_{VMC}=\left< \Phi_T\right|e^{-J}H e^{-J} \left|\Phi_{T}\right>
/(\left< \Phi_T\right|e^{-2J} \left|\Phi_{T}\right>)$, or the variance
of the ground state energy.  Replacing $K_{VMC}$ into Eq.
(\ref{eq:vK}) leads to a procedure similar to the optimization of
Filippi and Fahy \cite{filippi00} providing additional support to that
method. In the case of Refs. $\cite{wagner03,kolorenc}$ the nodes are
selected by adjusting the mix of density functionals that gives the
lowest DMC energy for a small system. The same mix is then used in a
larger system.  This procedure is in fact equivalent to optimizing the
shape of the effective potential with the restriction of remaining a
linear combination of two of more exchange-correlation potentials.
 We find that the change in the effective potential
required to optimize the nodes could be of the order of the {\it
  Hartree} potential, since the wave-function with the best nodes is
the non-interacting ground state, without any effective potential, for
all the range of interaction strengths and shapes explored.  Our
results suggest that counter intuitive directions for potential
optimizations should be explored to improve the nodes.

Potential optimization has also been applied for the prediction of
electronic excitations. Since $\rho({\bf r})$ determines $V({\bf r}) $
(but from a constant), the excitation spectra $\{E_{\nu,n} \}$ is a
function of $\rho({\bf r})$. This allows defining cost functions
$K_{ex}$ to match the spectra of a non-interacting system. In order to
minimize $K_{ex}$ one should do the derivatives $\delta
\varepsilon_{\nu } / \delta V_{KS}({\bf r})$ as in Ref \cite{us}.
When $\{E_{\nu} \}$ is taken from experiment, the search of a
potential giving a non-interacting density that minimizes $K_{ex}$ is
equivalent to the empirical potential method.\cite{wang95}.
Unfortunately, in this case the electronic density can no longer be used to obtain
the forces on the atoms. The existence of a single density functional
that can be used to obtain the excitation spectra of any system is
then a subject of debate. 

In summary, although the popular languages of electronic structure
theory all share the same quantum mechanical underpinnings, when
applied by experts to physical systems we often reach different
conclusions.  Many experts in QMC prefer HF wave-functions, while in
contrast calculations done within the GW-BSE approach often rely on
LDA derived wave-functions and energies\cite{aulbur00}, while some
hybrid density functionals obtain single particle excitations in
direct agreement with excitation spectra\cite{barone05}.  We argue
that as different theories need to retain {\it different} properties
of the {\it same ground state} wave-function to minimize errors,
different functionals should also be used. In some cases these
functionals correspond to the minimization of cost functions designed
to retain properties of the many body ground state in the
non-interacting wave-function.  Since some properties are favored at
the expense of others, it is unlikely that we can use the same
functional universally: we find that a function designed for optimal
nodes is a bad source of densities and vise versa. Here we give a
qualitative picture of the size of the differences that one can expect
as correlations start to dominate. With increasing interaction
strength, the exact Kohn-Sham non-interacting wave-function becomes a
much poorer description of several properties of the many-body ground
state. Methods that go beyond DFT are limited to the nearly
non-interacting limit if they depend strongly on DFT derived
wave-functions.

Research performed at the Materials Science and Technology Division
and the Center of Nanophase Material Sciences at Oak Ridge National
Laboratory sponsored the Division of Materials Sciences and the
Division of Scientific User Facilities U.S. Department of Energy. The
authors would like thank R.  Q. Hood, M. Kalos and M.  L. Tiago for
discussions.


%


\begin{thebibliography}{99}
\bibitem {practice} In practice, an approximate exchange
correlation potential is used.  Due to this approximation
$\bar\rho(r)$ and $\rho(r)$ are in practice different.
\bibitem{units} We define the energy unit to be $\hbar^2/(2 m \pi^2)$.
\bibitem{notebook} The Mathematica notebook is available upon request.
\end{thebibliography}
\end{document}